\documentclass[a4paper]{article}

\usepackage[UKenglish]{babel}
\usepackage[utf8]{inputenc}
\usepackage[T1]{fontenc}

\usepackage[hyphens]{url}

\usepackage[autostyle=true]{csquotes} 
\usepackage[backend=biber,style=alphabetic,maxnames=10,abbreviate=true,ibidtracker]{biblatex}

\usepackage[most]{tcolorbox}
\providecommand{\tightlist}{%
	\setlength{\itemsep}{0pt}\setlength{\parskip}{0pt}} 

\bibliography{references}

\begin{document}
	\title{Perceptions of YouTube's political influence}
	\date{1 April 2020}
	\author{Yury Kolotaev\footnote{Yury is a Master’s student in International Relations at Saint Petersburg State University. \texttt{firstname.lastname@mail.ru}}\,\, and Konrad Kollnig\footnote{Konrad is a student at Hertford College, Oxford. In his free time, he's interested in YouTube's influence on elections.
	\texttt{firstnamelastname@web.de}.}}
	\maketitle
	
    \begin{abstract}
        YouTube plays an ever more important role as a political medium.
        Yet, the implications are to-date not well understood and difficult to analyse, since access to YouTube's statistics is limited.
        To address this gap, we surveyed 124 people about their views and experiences around YouTube's political influence.
        Our results revealed diverse, sometimes conflicting views on YouTube's growing political role, and highlight the need for more research, discussion and possibly regulation.
    \end{abstract}
	

	\section{Introduction}
    
	The ongoing shift in the consumption and distribution of information brings rapid changes to politics.
	One of the main sources of transformation comes from social media, providing innovative ways to connect people worldwide.
	Unfortunately, most social media platforms, including YouTube, do not provide researchers with access to meaningful statistics about their platforms.
	The influence of these platforms on politics, and society as a whole, is hence not well understood, despite an ongoing public debate and high societal relevance.
    
    YouTube is the second-most visited website, after Google, and also the social network with the second-highest number of users, after Facebook\footfullcite{webstats}. This makes the platform one of the most important (online) media.
    As such, YouTube is becoming an essential medium for politics, to connect politicians with their audiences, inform citizens, and conduct electoral campaigns.
    
    
    
    The political influence of YouTube is yet widely underanalysed.
    Being a proprietary system, such analysis is difficult and not supported by the YouTube owner, Google.
    Consequently, the analyses to-date tend to revolve around anecdotal evidence about this subject, whilst quantitative insights are rare.
    
    The authors previously addressed questions and problems around YouTube's growing political role in an article in the \textit{Oxford Student}\footfullcite{oxford_student}, a student newspaper.
    As a continuation of the ideas and findings set out in this article, we surveyed $124$ people on their perception of YouTube's political influence.
    This is to gain some understanding of how YouTube influences politics, even without access to the internal statistics of YouTube.
	
	\section{Methodology}
	
	\begin{figure}
   	\begin{tcolorbox}
   		\begin{enumerate}
   			\tightlist
   			\item \textbf{YouTube usage}
   			\begin{enumerate}
   				\tightlist
   				\item How often do you watch YouTube videos?
   				\item Do you ever doublecheck information from YouTube?
   				\item What [problematic content] have you experienced on YouTube?
   			\end{enumerate}
   			\item \textbf{YouTube and politics}
   			\begin{enumerate}
   				\tightlist
   				\item How often do you consume information about politics?
   				\item How important is YouTube for you to learn about politics?
   				\item YouTube is a legitimate tool for political campaigning. Do you agree?
   				\item Why do you feel like this? (free-text question)
   				\item Is a YouTube channel a necessity for a politician today?
   			\end{enumerate}
   			\item \textbf{Video content on YouTube}
   			\begin{enumerate}
   				\tightlist
   				\item The \enquote{Up next} video recommendations on YouTube are helpful to see the videos that I want to see. Do you agree?
   				\item Would you prefer personalised video recommendations (based on your previous videos) or exploring new views on diverse topics?
   				\item Should there be more education about web platform usage and online misinformation?
   				\item Which type of content should be banned from YouTube?
   			\end{enumerate}
   		\end{enumerate}
   	\end{tcolorbox}
   	\caption{Interview questions and their three categories, excluding demographics.}
   	\label{fig:questions}
\end{figure}
	
	Our survey first asked for demographic details, and then 12 questions from three different categories: 1. \enquote{YouTube usage}, 2. \enquote{YouTube and politics}, and 3. \enquote{Video content on YouTube}.
	
	The first category of questions aimed to introduce the respondent to the topic, and understand their YouTube usage.
	The second category shed a light on the respondent’s awareness of the political dimension of YouTube.
	The last category considered the respondent’s understanding of the algorithmic personalisation of YouTube.
	These questions can be found in Figure~\ref{fig:questions}.
	
	11 questions were checkbox-style;
	7 of these employed \textit{Likert} questions (scale 1 to 5), which allow for quantitative evaluation.
	One question was free-text, asking to justify whether \enquote{YouTube is a legitimate tool for political campaigning}.
	
	\paragraph{Recruitment.}	
	The participants were mostly recruited through an article in an Oxford student newspaper\footfullcite{oxford_student}, that discussed recent examples of YouTube’s political influence and called for participation in our survey.
	This article and the survey were shared on social media (Facebook, Twitter and VK), mainly in Oxford and St Petersburg student groups.
	The survey was conducted over 10 weeks, from October 2019 to January 2020.
	The only requirement for participation in the survey was familiarity with YouTube.
	There was no reward for participation.
	
	\paragraph{Data analysis.}	
	The survey was checked for personally identifiable tokens, which were redacted for anonymity.
	The survey data was then analysed using Microsoft Excel, R, and Tableau for patterns within the quantitative data.
	In particular, we computed correlations and conducted $t$-tests for the 7 \textit{Likert} scale questions (on a scale 1 to 5).
	The free-text responses were first assigned to preliminary themes, then these themes were consolidated, and representative excerpts were chosen.
	
	\section{Results}
	
	A total of 124 persons (66 female, 55 male, 3 other) participated in the survey. The participants lived in 9 different countries, most of them in Russia (53\%), Germany (36\%) and the United Kingdom (5\%).
	The age ranges were: Under 18 (19 times), 18--20 (31 times), 21--24 (63 times), 25--29 (8 times), 30--39 (2 times) and 50--59 (1 time).
	
	In the following, the results for the three question categories are presented.
	Then, the \textsc{Likert} scale questions are evaluated.
	
	\begin{figure}
    \centering
    \includegraphics[width=\textwidth]{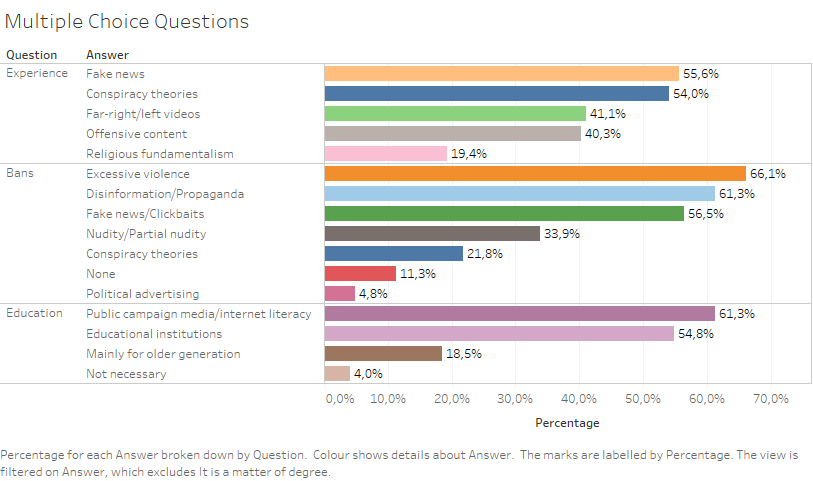}
    \caption{Results for the 3 multiple choice questions in the survey.}
    \label{fig:multiple_choice}
\end{figure}
	
\begin{figure}
 \centering
 \includegraphics[width=\textwidth]{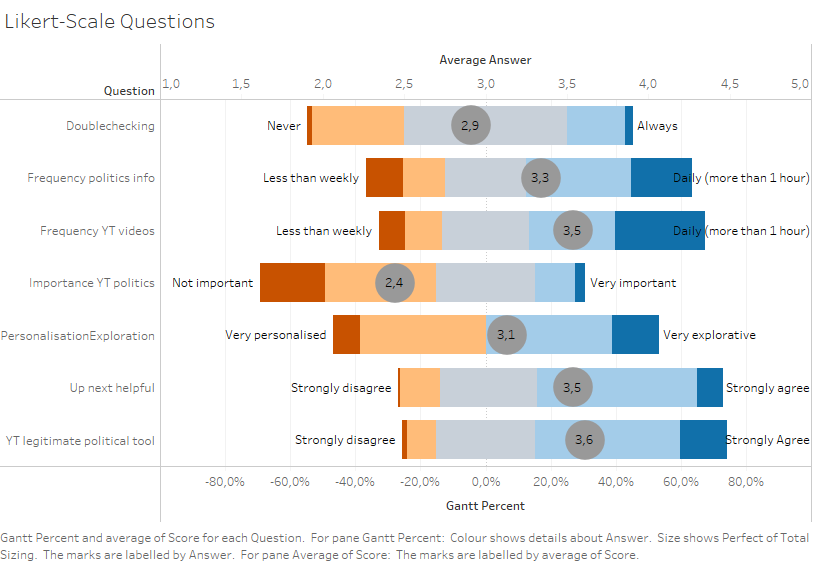}
 \caption{Visualisation of \textit{Likert}-scale answers as a \textit{Gantt} diagram. Shown are average scores, as well as the percentages of support for each answer (so-called \textit{Gantt} percentages).}
 \label{fig:likert}
\end{figure}

	\paragraph{YouTube usage.}	
	Most respondents watch YouTube \textit{daily} (27\% \textit{less than 1 hour}, 27\% \textit{more than 1 hour}), 27\% use YouTube \textit{multiple times per week}, and 11\% use it \textit{weekly}. 8\% stated to use YouTube \textit{less than weekly} (Variable \enquote{Frequency YT videos} in Figure~\ref{fig:likert}).

	Half of the participants (50\%) double-check information from YouTube \textit{sometimes}. 18\%, respectively 2\%, verify information \textit{very often}, respectively \textit{always}. 28\% rarely check YouTube information, and 2\% \textit{never} (Variable \enquote{Doublechecking} in Figure~\ref{fig:likert}).
	
	The respondents were asked what harmful content they had experienced on YouTube (Variable \enquote{Experience} in Figure~\ref{fig:multiple_choice}). 56\% of the participants have seen \textit{fake news}, 54\% \textit{conspiracy theories}, 41\% \textit{far-right/left} videos, 40\% \textit{offensive} content, and 19\% \textit{religious fundamentalism}. 20\% stated not to have seen any of these contents.
	
	\paragraph{YouTube and politics.}
	Within the second category of questions, the main focus was the political importance of YouTube.
	The majority consumes political content \textit{daily} (32\% \textit{less than 1 hour}, 19\% \textit{more than 1 hour}), 25\% \textit{multiple times per week}, 13\% \textit{weekly}, and 11\% \textit{less than weekly} (Variable \enquote{Frequency politics info} in Figure~\ref{fig:likert}).
	
	Most respondents consider YouTube a somewhat important source of information about politics: 3\% \textit{very important}, 12\% \textit{important}, 1\% \textit{moderately important}, 34\% \textit{slightly important}, and 20\% \textit{not important} (Variable \enquote{Importance YT politics} in Figure~\ref{fig:likert}).
	
	Two questions asked what place YouTube should take in politics.
	First, most respondents consider YouTube a legitimate tool for political campaigning. 44\% \textit{agree} and 15\% \textit{agree strongly} with the use of YouTube for politics. 31\% are \textit{neutral}, whilst 9\%, respectively 2\%, \textit{disagree}, respectively \textit{strongly disagree} (Variable \enquote{YT legitimate political tool} in Figure~\ref{fig:likert}).
	
	59 participants (48\%) provided a free-text justification for their choice.
	
	Second, a related question was whether a YouTube channel is a necessity for a politician today. The majority (63\%) state it \textit{depends on politician and electorate}. 11\% note the \textit{necessity} for every significant politician. At the same time, 21\% consider it \textit{unnecessary} and 5\% \textit{reject YouTube activity} of politicians.
	
	\begin{figure}
    \centering
    \includegraphics[width=\textwidth]{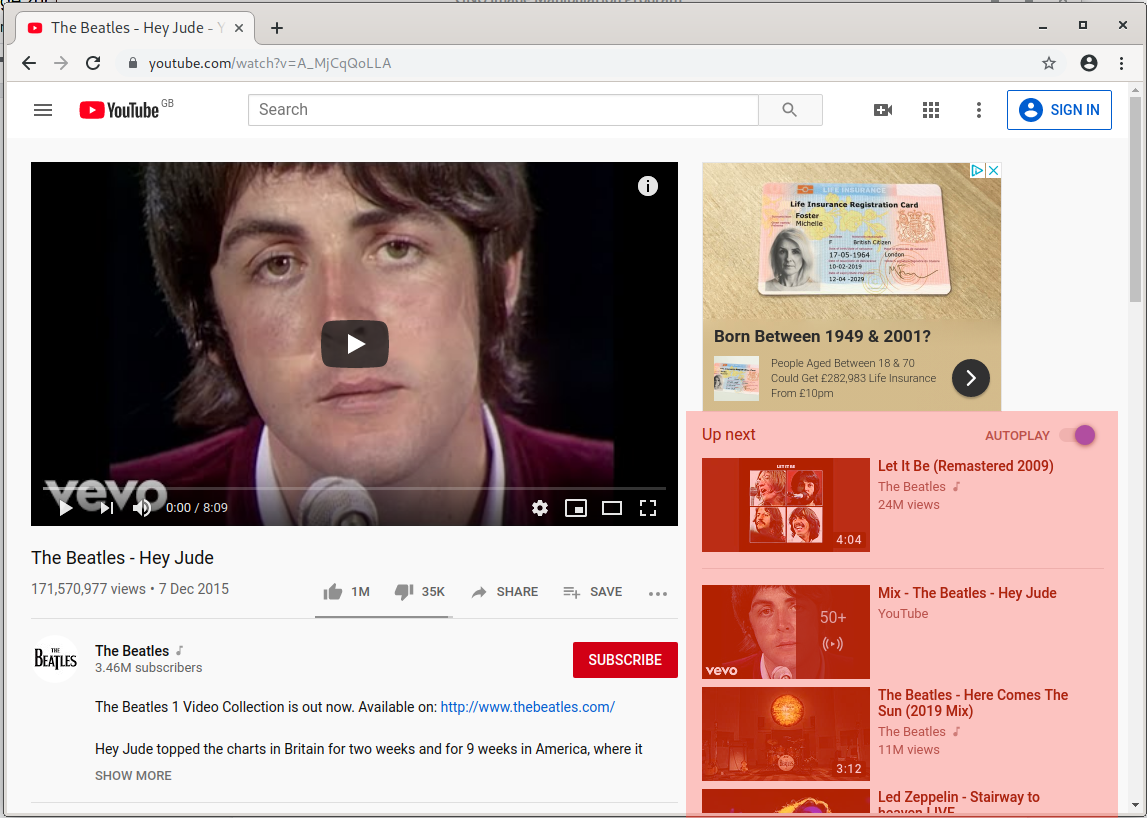}
    \caption{YouTube personalises the user experience, by suggesting videos in \enquote{Up next}.}
    \label{fig:upnext}
\end{figure}
	
	\paragraph{Video content on YouTube.}
	The last section covers the participants' experience with the YouTube personalisation algorithm.
	First, the participants were asked whether they find the \enquote{up next} recommendations helpful (see Figure~\ref{fig:upnext} and variable \enquote{Up next helpful} in Figure~\ref{fig:likert}).
	Most participants agree with the usefulness of the personalised section (49\% \textit{agree}, 8\% \textit{strongly agree}). 30\% were \textit{neutral}, whilst 12\% \textit{disagree} and 1\% \textit{strongly disagree}.
	
	Next, respondents could decide on a scale from 1 to 4 what they prefer: \textit{exploration and content diversification} or \textit{personalisation based on previous experience}.
	We found a slight tendency towards content exploration.
	15\% chose \enquote{4}, maximum content exploration, whilst 8\% chose \enquote{1}, maximum personalisation.
	The middle options \enquote{2} and \enquote{3} had equal support of 39\% each (Variable \enquote{PersonalisationExploration} in Figure~\ref{fig:likert}).
	
	The two last questions addressed how to tackle the current problems with YouTube.
	
	Participants were asked about the necessity of education about web platform usage and online misinformation (Variable \enquote{Education} in Figure~\ref{fig:multiple_choice}).
	61\% support \textit{a broad public campaign on media and internet literacy} and 55\% \textit{Internet literacy provision through educational institutions}.
	19\% stated that educational measures are mainly \textit{necessary for older generations} and 4\% consider such education on web platform usage as \textit{unnecessary}.
	
	Lastly, users could choose what type of content should be banned from YouTube (Variable \enquote{Bans} in Figure~\ref{fig:multiple_choice}).
	A majority selected \textit{excessive violence} (66\%), \textit{disinformation/propagand}a (61\%), \textit{fake news/clickbaits} (57\%).
	No majority supported bans of \textit{(partial) nudity} (34\%), \textit{conspiracy theories} (22\%), \textit{political advertising} (5\%).
	11\% of all participants \textit{do not promote bans} of any of the aforementioned content.
	
	{}
	
	\paragraph{Quantitative answers.}
	For the 7 \textsc{Likert} scale question, we computed the average answers, as well as the percentages of respondents for each answer, see Figure~\ref{fig:likert}.
	We also computed correlations between the variables.
	The three strongest correlations were between the following pairs of variables:
	\begin{itemize}
		\item (PersonalisationExploration, Up next helpful): $r=-0.41$, $p=0.000$,
		\item (Frequency YT videos, Importance YT politics): $r=0.30$, $p=0.001$, and
		\item (YT legitimate political tool, Frequency YT videos): $r=0.23$, $p=0.011$.
	\end{itemize}
	
	\section{Discussion}
	\subsection{YouTube, a legitimate tool for politics?}
	\label{sec:themes}
	Of the 124 respondents, 59 provided free-text answered about \textit{their view on the use of YouTube for politics}.
	These responses were a source of rich insights into the participants’ perceptions and motivated closer inspection of the occurring themes.
	
	We found through thematic analysis that the answers covered three main themes: \textit{Acceptance} (38 answers), \textit{Denial} (9 answers), and \textit{Indifference} (9 answers).
	Some answers mentioned several themes. There was even a group of participants who showed a \textit{sceptical attitude} to all of these themes (7 answers), and fell somewhere in between.

	\paragraph{Acceptance of YouTube as a political medium.}
	The first theme can be described as \textit{Acceptance} of YouTube becoming a widely used political tool.
	People expressing \textit{Acceptance} assume that YouTube has the same role as other audiovisual media, such as TV or radio.
	According to these participants, \textit{\enquote{it’s just another kind of media}}.
	YouTube often serves as a distribution channel for classical media and does not has to be regulated differently.
	
	Another reason to see YouTube as a legitimate platform for political advertising is higher freedom of speech than classical media, being more equal and less discriminatory.
	Compared to other media, YouTube's editorship of content is limited.
	The flow of information is rather in the hands of many than of few actors, and overall less subject to control.
	\textit{\enquote{there's more freedom for political activity on YouTube than at TV or during the offline activity}}.
	This can provide a serious reason to users to accept YouTube's expanding political influence.
	
	Some respondents even apply this view of uncontrolled information to advertising on YouTube.
	Content (whether political or not) \textit{\enquote{can be advertised just like beauty products can be}}.
	
	\paragraph{Indifference to YouTube’s political role.}
	The \textit{Indifference} theme shows no strong position on YouTube's political role, or sees no connection to politics.
	They state that \textit{\enquote{for many people YouTube is still a way to entertain, not to get important or official information}}.
	For them, YouTube is just a product of technological progress.
	
	In fact, some participants assume that official parties would not use YouTube for self-promotion.
	Others have never considered this before, or confess that they cannot make an informed judgement.
	
	\paragraph{Denial of YouTube’s legitimacy in politics.}
	The last theme opposes YouTube’s legitimacy as a political instrument, and laments the lack of quality and control of information on YouTube.
	One of the main arguments stressed by this group is the \textit{\enquote{bias in the YouTube recommendation algorithm}} that changes political campaigning in an undemocratic way.
	These algorithms can make the presence of a certain political idea or position strongly linked to the financial means of a politician.
	This can, in return, lead to \textit{\enquote{injustice in competition}}, turning YouTube advertising into an unfair tool in the sense of pluralism of political content and representation.
	
	Overall, this group of participants perceives YouTube as \textit{\enquote{not neutral}}.
	The respondents emphasised that YouTube \textit{\enquote{isn’t qualified enough to be a real source of politics content}}.
	There is a large amount of unchecked information on YouTube, without editing. As such, YouTube facilitates an uncontrolled information flow, some of which is potentially harmful.
	This makes YouTube different to traditional media, that have quality control and editorial liability.
	Traditional media limits the dissemination of biased, false, or deceptive information or content provided by \textit{\enquote{unprofessional}} creators better than YouTube.
	
	\subsection{YouTube, a platform or an edited medium?}
	The above themes stress the importance of the legal question of whether YouTube is a publisher, i.e. an edited medium, or a platform.
	If YouTube is just a platform for content sharing, then the YouTube owners would not be liable for any content shared.
	
	If, however, YouTube is an edited online medium, YouTube would face similar liabilities as online media.
	YouTube could potentially be liable for \enquote{\enquote{incitement} or \enquote{negligent publication} if a reader of their publication is seriously injured}\footfullcite{findlaw}.
	
	The core of these questions revolves around whether YouTube’s content selection algorithms provide editorship of the content displayed.
	Opponents argue that YouTube’s scale necessitates the use of algorithms, and there are no realistic means through which the operation of the platform could be carried out otherwise.
	There is no human editorship, and hence there can be no liability. Further, any explicit human intervention would constitute censorship, and potentially infringe basic human rights, such as freedom of speech.
	Others disagree, noting that the development of YouTube’s algorithms alone requires a high level of human engineering.
	By doing so, the owners of YouTube can indeed influence the content seen and the values propagated  \cite{Gillespie2010,Zuboff2018}.
	
	This discussion goes against the foundation of YouTube’s business model, that profits from keeping users as long as possible on the platform, to show them more advertising. In this view, the value of content is mainly determined by the \enquote{engagement} of users.
	Research has shown that inflammatory material is indeed more \enquote{engaging} for users, appealing to their emotions. Fact-checked content is usually less emotional, and hence less engaging and valuable for YouTube \cite{Vosoughi2018}.
	
	The identified theme \textit{Acceptance} supports YouTube's libertarian approach and is, in fact, expressed in the majority of free-text answers (64\%).
	So, users seem to support YouTube's business model as is, or do not see an alternative.
	This is supported by our quantitative results, discussed in the following.
	
	\subsection{Users like YouTube, particularly the young}
	The six \textsc{Likert} scale questions of our survey provide further quantitative insights into respondents' perceptions.
	None of the correlations would be considered \enquote{strong} in the literature.
	However, they align with our expectations, and raise some worrying concerns.
	Hence, we discuss the three strongest correlations, to expand our discussion and spur further research.
	
	\paragraph{Users like YouTube’s suggestions.}
	We found that the more a respondent preferred content similar to previous videos, rather than new views and diverse content, the more they found YouTube’s up next recommendations helpful.
	This could be support of respondents to staying in their filter bubble, which some researchers argue deepens societal divide.
	
	Yet, these results might be influenced by the fact that our survey asked participants about negative experiences with YouTube, immediately before.
	This would then reflect a user's attitude to avoid negative surprises, also a characterising feature of the filter bubble.
	
	\paragraph{YouTube use and political acceptance.}
	We found two further interesting, weak correlations with the frequency of YouTube use.
	The more often a respondent uses YouTube,
	\begin{enumerate}
	    \item the more important they found YouTube for politics, and
	    \item the more legitimate they found YouTube as a political tool.
	\end{enumerate}
	For instance, 64 \% of those who use YouTube daily agree that YouTube is a legitimate tool for political campaigning.
	This correlation is supported by the fact that 71\% of the free-text answers, expressing \textit{Acceptance} of the political use of YouTube, stem from these daily users, whilst only accounting for 55\% of the overall respondents.
	
	\paragraph{Age and YouTube use.}
	We analysed the relation between participant age and the frequency of YouTube usage.
	Younger participants showed to spend more time on YouTube.
	
	We noticed that 37\% of the respondents under 18 watch YouTube for more than one hour daily, whilst only 32\% participants aged 18--20 do, and 24\% of those 21--24.
	Such results are expected, since the younger generation tends to be more digitally apt than their older peers.
	Despite such frequent use of YouTube, the level of verification and double-checking of YouTube content is only moderate amongst the participants of our study.
	
	Yet, these results are unsettling, because young people are more susceptible to manipulation \cite{mcgrew2017challenge}. They are less in the position to distance themselves from content seen, distinguish between what is real and what is not, and critically reflect upon others' opinions.
	
	Overall, this reduced agency of the young requires special safeguards, and may pose a high risk to individual development and society as a whole.
	
	\subsection{Inconsistent answers}
	\label{sec:inconsistencies}
	We found that the answers of some participants were sometimes inconsistent, thus motivating further reflection.
	The most striking example is the contradiction in the experienced content on YouTube, and the opinions what kind of content should be banned from YouTube.
	
	The vast majority of most participants have experienced harmful content, such as misinformation or violence, except 
	24 respondents (19\%).
	Yet, 22 of these respondents (92\%) propose to ban certain harmful content from YouTube.
	The emerging question is how can someone support limitations of content without having experienced it?
	Leaving aside the limitations of the study, there are at least three explanations. These explanations are not distinct, and they might overlap.
	\begin{enumerate}
	    \item \textbf{Values.}
	    The respondent bases his judgement on his normative values, and rejects any harmful content. 
	    By doing so, the respondent implicitly weighs the harm of such content higher than freedom of speech. With such an attitude the real experience of harmful content is irrelevant. Values prevail.
		\item \textbf{Securitisation.}
		The idea of possible bans or restrictions online is currently widely discussed.
		Such a politicisation of fake news can lead to a societal opinion that any limitation or even prohibition of harmful content is perceived as inevitable and necessary.
		This phenomenon of securitisation is well known in political science and could have contributed to the participant's response.
		\item \textbf{Unconsciousness.}
		It is possible that participants did experience harmful content, such as fake news, but did not notice.
		Such fake content might fit their filter bubble well, in the absence of correcting and conflicting narratives online.
	\end{enumerate}
	This third explanation, unconsciousness, is the most dangerous and alarming one.
	If users cannot make an informed decision about their YouTube use, we may have to think about ways to enable them to do.
	
	Overall, inconsistencies in responses underline that the perception of YouTube is subject to cognitive biases and limitations, and this may indeed require intervention.
	
	
	
	
	\subsection{Limitations}
	
	This study comes with several limitations.
	The number of participants (124) is very small for a survey, and the participants are not representative, being mostly students.
	In fact, participants come from various different countries (mostly Russia, Germany, and the UK), with possibly different perceptions of YouTube.
	
	Moreover, the recruitment through a newspaper article on political advertising on YouTube will have further selected and primed respondents.
	We chose this approach to raise awareness for the under-studied topic of YouTube's influence, and reach a large audience without monetary expenses.
	
	Further, surveys come with profound limitations and responses never fully reflect reality.
	Users merely report on their perceived experience.
	In doing so, users are often inconsistent in their behaviour, as highlighted in Section~\ref{sec:inconsistencies}.
	We applied standard mitigation strategies, such as motivational questions, to improve answer quality.
	
	
	
	\section{Conclusions}
	This report highlighted the diverse, sometimes conflicting views on YouTube as an emerging political instrument and medium.
	
	We found that a vast majority of respondents both accepts the emerging political role of YouTube, and supports bans of harmful content.
	Yet, imposing such bans is difficult, especially when some content serves a political role.
	The nearly insoluble task is to draw the line, without unfairly restricting political participation.
	
	These observation highlight the wider conflict of whether YouTube classifies as a platform or a publisher.
	If it were a publisher, YouTube would have to face similar liabilities over its content and editorship as classical media.
	Even if YouTube continues to exist as a (political) platform, the question of how to treat harmful content will still persist and require further discussion.

    We found that YouTube as a medium is particularly important for young users, who spend significantly more time on YouTube than their older peers.
    Those young users tended to express less concern about YouTube's growing political role.
    Yet, the young are known to be particularly susceptible to manipulation \cite{mcgrew2017challenge}.
    Special safeguards are required, which the platform currently fails to implement.
    
    Overall, our observations give reason to reflect on the necessary changes to political processes in the digital age and the role YouTube should play in this context.
    \printbibliography
	
\end{document}